\documentclass[aps,preprintnumbers,amsmath,amssymb,prd,superscriptaddress,longbibliography,twocolumn]{revtex4-2}
\usepackage{dcolumn}% Align table columns on decimal point
\usepackage{color}
\usepackage[caption=false]{subfig}
\usepackage{feynmp}
\usepackage{feynmp-auto}
\usepackage{float}
\usepackage{bbm}
\usepackage{empheq}
\usepackage{amssymb}

\def\comment#1{}

\newcommand{\nc}{\newcommand}  

\nc{\scs}{\scriptstyle}
\nc{\setval}{\fmfset{wiggly_len}{3mm} \fmfset{arrow_len}{1.5mm}
	\fmfset{arrow_ang}{13} \fmfset{dash_len}{1.5mm}\fmfpen{0.125mm}
	\fmfset{dot_size}{2thick}}

\usepackage{bm,latexsym,mathrsfs,enumerate,color}
\usepackage[mathcal]{euscript}
\usepackage[breaklinks=true,unicode=true,urlcolor = blue,colorlinks = true,citecolor = blue,linkcolor = blue]{hyperref}
\usepackage{graphicx}
\usepackage{todonotes}
\usepackage{wrapfig}

\renewcommand{\vec}[1]{\bm{#1}}

\def\slashchar#1{\setbox0=\hbox{$#1$}           % set a box for #1
	\dimen0=\wd0                                 % and get its size
	\setbox1=\hbox{/} \dimen1=\wd1               % get size of /
	\ifdim\dimen0>\dimen1                        % #1 is bigger
	\rlap{\hbox to \dimen0{\hfil/\hfil}}      % so center / in box
	#1                                        % and print #1
	\else                                        % / is bigger
	\rlap{\hbox to \dimen1{\hfil$#1$\hfil}}   % so center #1
	/                                         % and print /
	\fi}                                         %

\DeclareMathAlphabet\mathbfcal{OMS}{cmsy}{b}{n}

\usepackage{physics}
\usepackage{xcolor}

\begin{document}

%\preprint{APS/123-QED}

\title{Antiferromagnetism and spin excitations in a two-dimensional non-Hermitian Hatano-Nelson flux model}

\author{Eduard Naichuk}
%\email[Correspondence to: ]{eduardnaich@gmail.com}
\affiliation{Institute for Theoretical Solid State Physics, IFW Dresden, Helmholtzstr. 20, 01069 Dresden, Germany}
\affiliation{Bogolyubov Institute for Theoretical Physics, 03143 Kyiv, Ukraine}

\author{Ilya M. Eremin}
\affiliation{Theoretische Physik III, Ruhr-Universit\"at Bochum, D-44801 Bochum, Germany}

\author{Jeroen van den Brink}
\affiliation{Institute for Theoretical Solid State Physics, IFW Dresden, Helmholtzstr. 20, 01069 Dresden, Germany}
\affiliation{Institute for Theoretical Physics and W\"urzburg-Dresden Cluster of Excellence ct.qmat, TU Dresden, 01069 Dresden, Germany}

\author{Flavio S. Nogueira}
\affiliation{Institute for Theoretical Solid State Physics, IFW Dresden, Helmholtzstr. 20, 01069 Dresden, Germany}

\date{\today}% It is always \today, today,
             %  but any date may be explicitly specified

\begin{abstract}
The one-dimensional Hatano-Nelson model with non-reciprocal hoppings is a prominent example of a relatively simple non-Hermitian quantum-mechanical system, which allows to study various phenomena in open quantum systems  without adding extra gain and loss terms. Here we propose to use it as a building block to construct a correlated non-Hermitian Hamiltonian in two dimensions. It has the characteristic form of a flux model with clock-anticlockwise non-reciprocal hopping on each plaquette. Adding the on-site Hubbard type interaction we analyze the formation of the longe-range antiferromagnetic order and its spin excitations. Such a model is non-Hermitian, but $\mathcal{PT}$-symmetric, which leads to the existence of two regions: a region of unbroken $\mathcal{PT}$ symmetry (real-valued spectrum) and a region of broken $\mathcal{PT}$ symmetry with exceptional lines and complex-valued energy spectrum. The transition from one region to another is controlled by the value of the on-site interaction parameter and coincides with the metal-insulator transition. We also analyze the spin wave spectrum, which is characterized by two diffusive $d$-wave type of modes corresponding to gain and loss.  
\end{abstract}

%\keywords{Suggested keywords}%Use showkeys class option if keyword
%display desired
\maketitle

%\tableofcontents

\section{Introduction}

In recent years one of the most popular avenues to study the evolution of open quantum-mechanical systems became its description via effective non-Hermitian Hamiltonians, which break unitarity \cite{Ashida_2020}. At the origin of these research activities lies the seminal work on parity-time ($\mathcal{PT}$) symmetric Hamiltonian with balanced dissipation (loss) and gain energy terms added to Hermitian parent Hamiltonians \cite{Bender_1998,Bender_2005}. $\mathcal{PT}$ symmetry allows to construct further non-Hermitian Hamiltonians, which for some parameter space yield real eigenvalues and the transition between purely real and complex-valued (dissipative) energy spectrum is separated by exceptional points \cite{ElGanainy2018,Kawabata2019}. The field of non-Hermitian systems has grown in recent years, driven by theoretical predictions and experimental discoveries across various field of physics such as optics, acoustics, ultra-cold atomic gases, and superconducting qubits \cite{Ashida_2020,Ding2022,Okuma2023}. In particular, non-Hermitian Hamiltonians have recently become quite promising due to the possibility of different realizations, for example using ultracold atoms \cite{Mu, Liang}, photonics \cite{Weidemann}, metamaterials \cite{Brandenbourger2019, Aoxi_Wang} and topological electrical circuits \cite{Helbig, Stegmaier, Liu}. The characteristic physical phenomena of non-Hermitian systems, such as formation of exceptional points, chiral transport around them, and the non-Hermitian skin effect are typically understood within a single-particle Hamiltonian formalism, and currently the interest shifts towards many-body systems where the interplay between interactions and non-Hermiticity plays a central role. 

\begin{figure*}
    \includegraphics[width=1.0\linewidth]{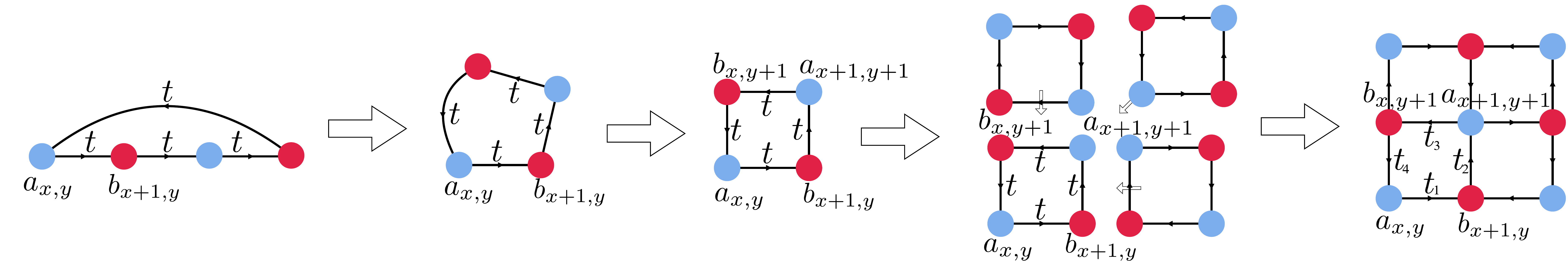}
 	\caption{The illustration of creating two-dimensional version of the HN model using completely asymmetric one-dimensional Hatano-Nelson model with periodic boundary conditions by constructing plaquettes with four sites.} 
 	\label{Fig:Hopping}
\end{figure*}

One prominent example of non-Hermiticity appears in the celebrated Hatano-Nelson (HN) model \cite{Hatano_Nelson_1996, Hatano_Nelson_1997, Hatano_Nelson_1998}, which consists of a one-dimensional (1D) lattice with asymmetric (non-reciprocal) hoppings, responsible for the
localization of all the bulk states at the edges, which leads to
the emergence of the so-called topological non-Hermitian skin effect. The non-reciprocity of the hopping may be considered as resulting from some transverse pseudomagnetic field in a cylinder geometry \cite{Hatano_Nelson_1996}. By now, there has been studies of this model in the presence of interactions \cite{Zhang2022},  disorder {\color{blue}\cite{Hatano_Nelson_1996,Kim2001,Hamazaki2019,Orito2023,Mk2024}}, as well as its extension to two-dimensional lattices \cite{Edvardsson2022,Hou2024}. Motivated by the HN model, we propose a realization of non-Hermiticity in two dimensions exploring a certain resemblance with the Hermitian flux $t$-$\varphi$ model, where $\varphi$ is a staggered magnetic flux \cite{Harris}. Typically flux phases are constructed by considering hopping matrix elements having the form $t_{\vec{r}_1,\vec{r}_2}=|t_{\vec{r}_1,\vec{r}_2}|e^{i\varphi_{\vec{r}_1,\vec{r}_2}}$. For a Hermitian system we require $\varphi_{\vec{r}_1,\vec{r}_2}=-\varphi_{\vec{r}_2,\vec{r}_1}$. If the latter condition does not hold, we achieve a non-reciprocal hopping akin to the HN model. Our construction starts with a one-dimensional HN model with four sites and periodic boundary conditions having a maximally asymmetric hopping, i.e.,  one occurring  only in one direction, see  Fig.~\ref{Fig:Hopping}. Such a simple one-dimensional model is used as a building block for a two-dimensionional non-Hermitian system where each plaquette represents the maximally asymmetric HN model just described. The procedure is schematically illustrated in Fig.~\ref{Fig:Hopping}. By combining these plaquettes in a clock-anticlockwise fashion we arrive at the non-Hermitian version of the flux-like model. Note that the Hermitian $t$-$\varphi$ flux model includes gauge-invariant couplings to the magnetic field, and this can lead to an array of currents giving rise to an antiferromagnetic (AF) flux lattice \cite{Harris}, whereas our model is characterized by asymmetrical hopping and does not include gauge-invariant couplings to an actual magnetic field. It is also interesting to observe that the current pattern following from Fig.~\ref{Fig:Hopping} bears a resemblance to ice models, more specifically, in the form of a two-dimensional vertex model, which from the figure can be identified to the so called model F \cite{baxter2007exactly}. Also worth mentioning in this context is the vertex model associated to the XXZ Heisenberg chain \cite{McCoy-Wu-1968}, which also takes a non-Hermitian form.    

Having formulated the generalization of the HN model to two dimensions in the form of a flux-like model, it is tempting to consider the effect of the interaction in the system by adding an on-site Coulomb repulsion. It is known that at half-filling the system is prone towards antiferromagnetic Mott insulating phase, which then evolves towards metallic spin density wave (SDW) state at finite doping. Here, we analyze the formation of the long-range AF state in the proposed non-Hermitian model and its longitudinal and transverse spin excitations by computing the corresponding dynamical spin susceptibility. The structure of the paper is as follows. In Sec.~\ref{Model} we define the Hamiltonian of our model and study the behavior of the spectrum in different interaction regimes using the Hubbard-Stratonovich transformation. In Sec.~\ref{Dynamic_spin_susceptibility} we calculate dynamic spin susceptibilities for the transverse and the longitudinal components, and analyze our system for the presence of spin-waves. The spin-wave energy spectrum is found to be purely diffusive $d$-wave modes representing gain and loss. In Sec.~\ref{Conclusions_and_outlook} we summarize the main results and discuss prospects for further research.

\section{Model}
\label{Model}

\subsection{Definition of model}
The proposed generalization of the HN model in two dimensions in the form of the flux model has the following Hamiltonian,  
\begin{eqnarray}
	\label{H_NH}
	H&=&H_t-\mu\sum_{\boldsymbol{r},\sigma}f^{\dagger}_{\boldsymbol{r},\sigma}f_{\boldsymbol{r},\sigma}\nonumber\\&+&U\sum_{\boldsymbol{r}}\left(f^{\dagger}_{\boldsymbol{r},\uparrow}f_{\boldsymbol{r},\uparrow}-\frac{1}{2}\right)\left(f^{\dagger}_{\boldsymbol{r},\downarrow}f_{\boldsymbol{r},\downarrow}-\frac{1}{2}\right),
\end{eqnarray}
where,
\begin{eqnarray}
	H_t=-\frac{1}{4}\sum_{\vec{r},\sigma}\left(V_{\vec{r},\sigma}^A+V_{\vec{r},\sigma}^B\right),
\end{eqnarray}
with
\begin{equation}
    V_{\vec{r},\sigma}^A=t_4a_{\vec{r},\sigma}^\dagger(b_{\vec{r}+\hat{\vec{y}},\sigma}+b_{\vec{r}-\hat{\vec{y}},\sigma})+t_3(b_{\vec{r}+\hat{\vec{x}},\sigma}^\dagger+b_{\vec{r}-\hat{\vec{x}},\sigma}^\dagger)a_{\vec{r},\sigma},
\end{equation}
\begin{equation}
    V_{\vec{r},\sigma}^B=t_1b_{\vec{r},\sigma}^\dagger(a_{\vec{r}+\hat{\vec{x}},\sigma}+a_{\vec{r}-\hat{\vec{x}},\sigma})+t_2(a_{\vec{r}+\hat{\vec{y}},\sigma}^\dagger+a_{\vec{r}-\hat{\vec{y}},\sigma}^\dagger)b_{\vec{r},\sigma}. 
\end{equation}
The model is characterized by three parameters: the hopping integral $t$, chemical potential $\mu$, and the on-site Hubbard repulsion $U$. The operators $a^{\dagger}_{\boldsymbol{r},\sigma}$ and $b_{\boldsymbol{r},\sigma}$ are the creation and annihilation operators for fermions with spin $\sigma=\uparrow,\downarrow$ and site index $\boldsymbol{r}$ on the sub-lattices $A$ and $B$, respectively,
\begin{eqnarray}
	f_{\boldsymbol{r},\sigma}=\begin{cases}
	a_{\boldsymbol{r},\sigma}, \quad \boldsymbol{r}\in A\\
	b_{\boldsymbol{r},\sigma}, \quad \boldsymbol{r}\in B
	\end{cases}.
\end{eqnarray}

Although the Hamiltonian of Eq.~\eqref{H_NH} is non-Hermitian, it is $\mathcal{PT}$-symmetric. On a lattice, $\mathcal{P}$ and $\mathcal{T}$ operate as follows,  
\begin{eqnarray}
	\mathcal{P}f_{\boldsymbol{r},\sigma}\mathcal{P}=f_{\boldsymbol{N}+\boldsymbol{n}-\boldsymbol{r},\sigma}, \quad \mathcal{T}i\mathcal{T}=-i,
\end{eqnarray}
where $\boldsymbol{N}=(N,N)$, with $N$ being the number of lattice sites and $\boldsymbol{n}=(1,1)$.

\begin{figure}
    \includegraphics[width=8cm]{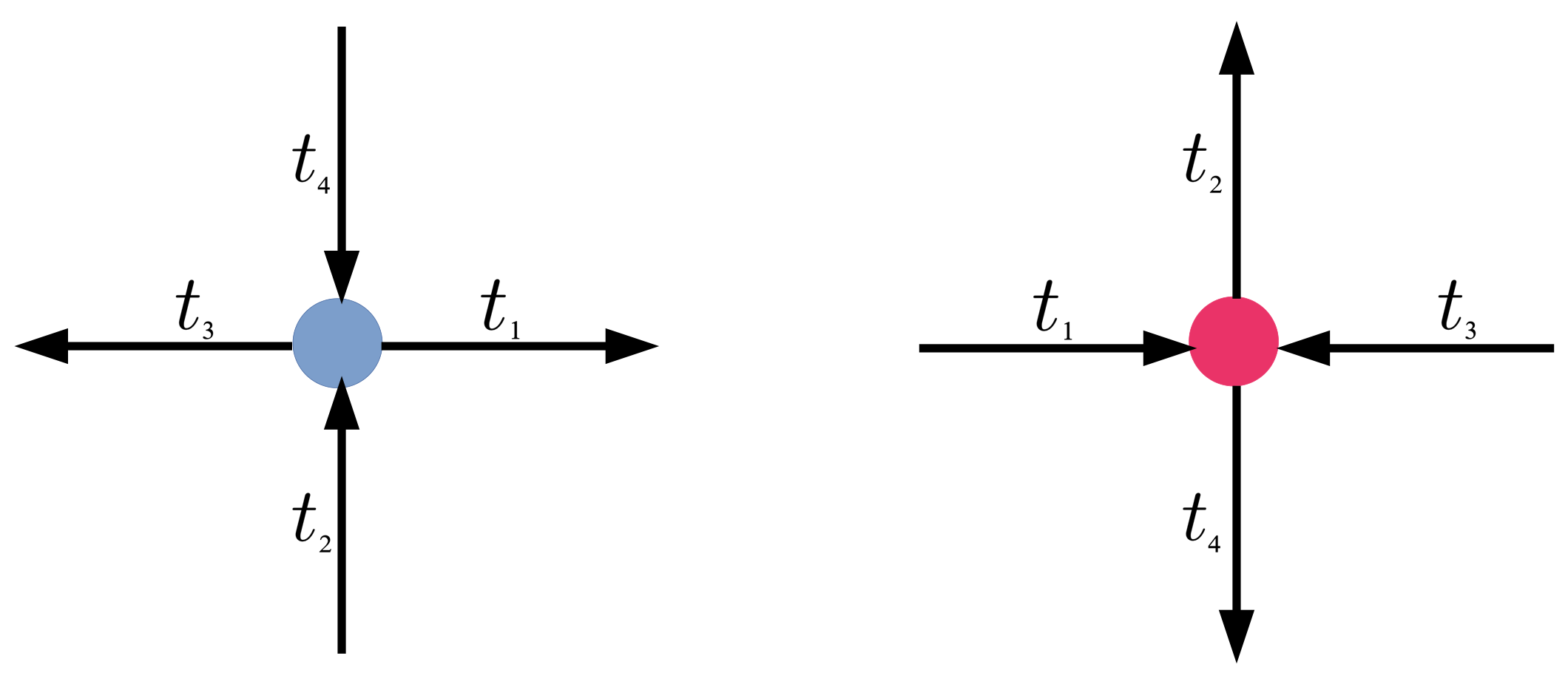}
    \caption{The two types of vertices appearing in the construction of the two-dimensional lattice. The directions of the arrows indicate the incoming and outgoing fermions. From this it is impossible to make a direct connection between vertices of the same type.}
    \label{Fig:Vertices}
\end{figure}

The hopping processes are shown in Fig.~\ref{Fig:Hopping}. In this case we can consider the two types of vertices occurring in the lattice of Fig.~\ref{Fig:Hopping}. The bipartite lattice is generated by the blue and red vertices shown in Fig.~\ref{Fig:Vertices}. Thus, picturially, the $\mathcal{PT}$-symmetry is realized by exchanging two vertices ($\mathcal{P}$) and inverting the direction of the arrows ($\mathcal{T}$). The two type of vertices are represented by the operators, $V_{\vec{r},\sigma}^A$ and $V_{\vec{r},\sigma}^B$.

\subsection{Model in the presence of a magnetic field}

Although this model does not include an external magnetic field, we can easily consider its effect by including the gauge-invariant couplings to an actual magnetic field using a Peierls substitution. In this case, the model should be characterized not only by a flux in terms of hopping, but also by a flux in terms of the external magnetic field. We can always choose a gauge in which the phase is equally distributed on all bonds so that $t_1=t_2=t_3^*=t_4^*=te^{-i\varphi}$, where $\varphi=\sum_{\langle\vec{r},\vec{r}'\rangle\in p}\varphi_{\vec{r},\vec{r}'}$ is the staggered magnetic flux with sum over all plaquettes $p$ and $\varphi_{\vec{r},\vec{r}'}=e/\hbar c\int_{\vec{r}}^{\vec{r}'}\vec{A}\cdot d\vec{l}$ is the phase, and $\vec{A}$ is the vector potential. It is easy to see that, the energy dispersion for $H_t$ is then given by $\mathcal{E}^{\pm}_{\vec{k}}=-\mu \pm \varepsilon_{\vec{k}}$, where 
\begin{equation}
\label{Eq:epsilon}
    \varepsilon_{\vec{k}}=t\cos{\varphi}\sqrt{\cos k_x \cos k_y}.
\end{equation}
Finally, the presence of a magnetic flux can be generalized by the substitution $t\to t\cos{\varphi}$. As a consequence, any calculation can be done as a function of a {\it gauge invariant} magnetic flux $\varphi$. However, for simplicity we will assume $\varphi=2\pi n$, $n\in\mathbb{Z}$ and concentrate on the properties of the model that with a staggered $2\pi$-flux (mod $n$). This differs of course from the staggered $\pi$-flux phase, but the non-Hermitian character of the system causes the $2\pi$ quantized flux to bear some similiraties with the latter, as we will see. 

%It is instructive first to look into the properties of the non-interacting model where $U=0$. It is easy to see that, the energy dispersion is given by  $\mathcal{E}^{\pm}_{\vec{k}}=-\mu \pm \varepsilon_{\vec{k}}$ where 
%
%\begin{equation}
%\label{Eq:epsilon}
%    \varepsilon_{\vec{k}}=t\sqrt{\cos k_x \cos k_y}.
%\end{equation}

\subsection{Properties of the spectrum}

\begin{figure}
    \subfloat[]{\includegraphics[width=0.49\linewidth]{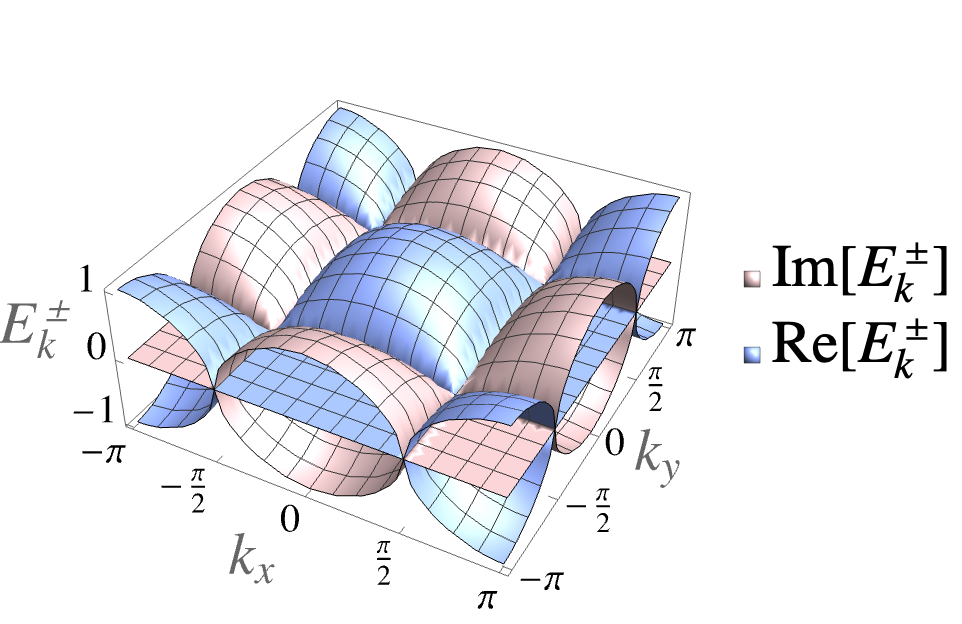}}
    \hspace{0.02cm}
    \subfloat[]{\includegraphics[width=0.49\linewidth]{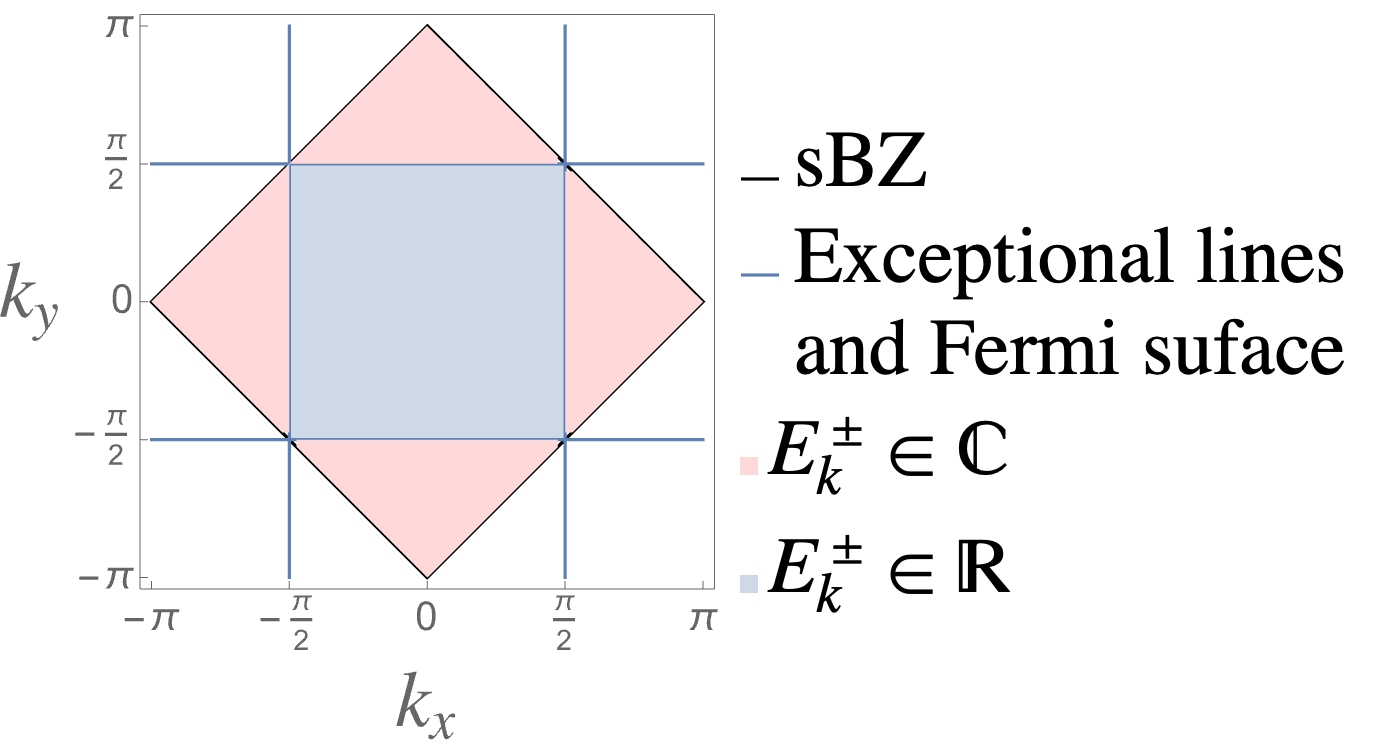}}
    \vfill
    \subfloat[]{\includegraphics[width=0.49\linewidth]{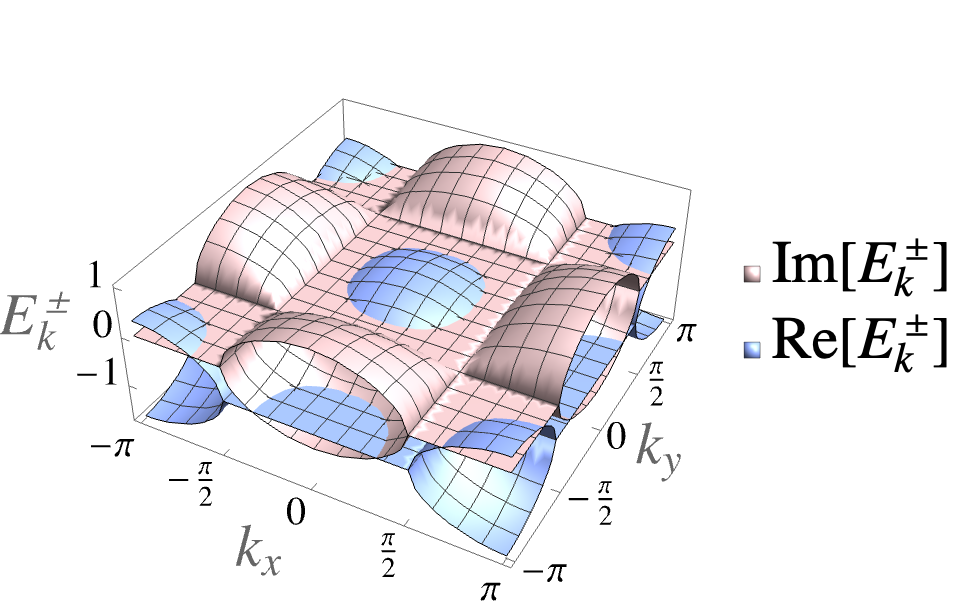}}
    \hspace{0.02cm}
    \subfloat[]{\includegraphics[width=0.49\linewidth]{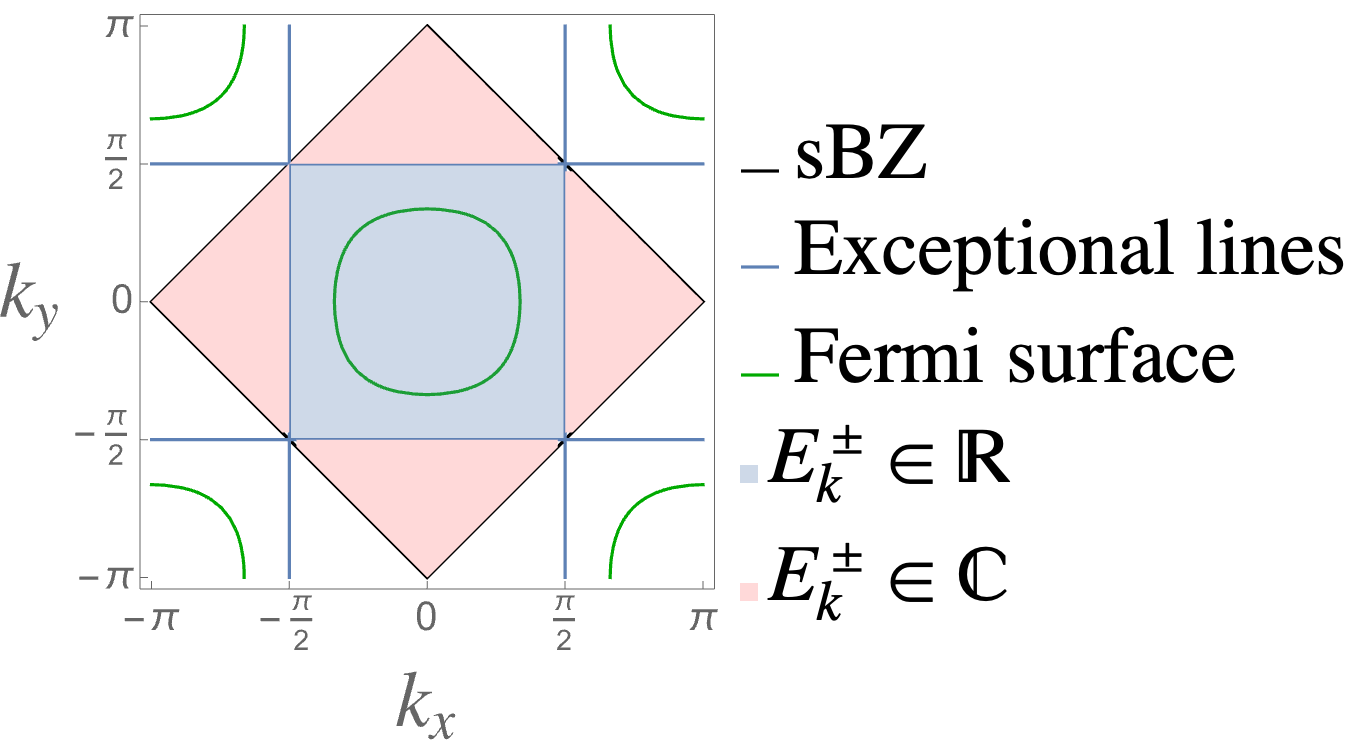}}
 	\caption{Calculated energy dispersion of the non-interacting ($W=0$) two-dimensional HN flux model for $\mu=0$ (a), $\mu=0.7t$ (c) in the first BZ. Panel (b) shows the Fermi surface $\mu=0$, which are also  exceptional lines $\cos k_x \cos k_y=0$. Panel (d) shows the Fermi surface for $\mu=0.7$ and  exceptional lines.} 
 	\label{Fig:Spectrum_W=0}
\end{figure}

For $U=0$ and $\mu=0$ (half-filling) the energy spectrum splits into real and imaginary parts, separated by exceptional lines determined by the condition $\cos k_x \cos k_y=0$. In particular, inside the sublattice Brillouin zone (sBZ), the spectrum is real in the distinct intervals $k_x\in(-\pi/2,\pi/2)$ and $k_y\in(-\pi/2,\pi/2)$ and there is no gap between the real and imaginary parts of the energies. Remarkably, the exceptional lines coincide here with the Fermi surface as shown in panels (a) and (b) of Fig.~\ref{Fig:Spectrum_W=0}. For finite $\mu \le t$ only one of the bands crosses the Fermi level, which is determined by $\mu = \mbox{Re}~\varepsilon_{\vec{k}}$ and it does not coincide with exceptional lines. For $\mu>t$ the system is a band insulator.  

\begin{figure}
    \subfloat[]{\includegraphics[width=0.49\linewidth]{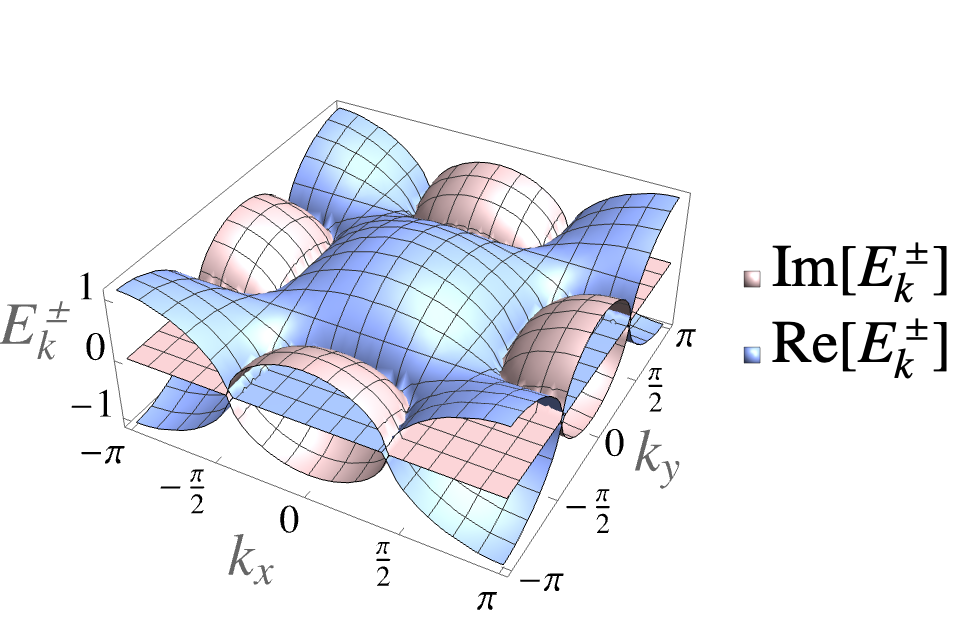}}
    \hspace{0.02cm}
    \subfloat[]
    {\includegraphics[width=0.49\linewidth]{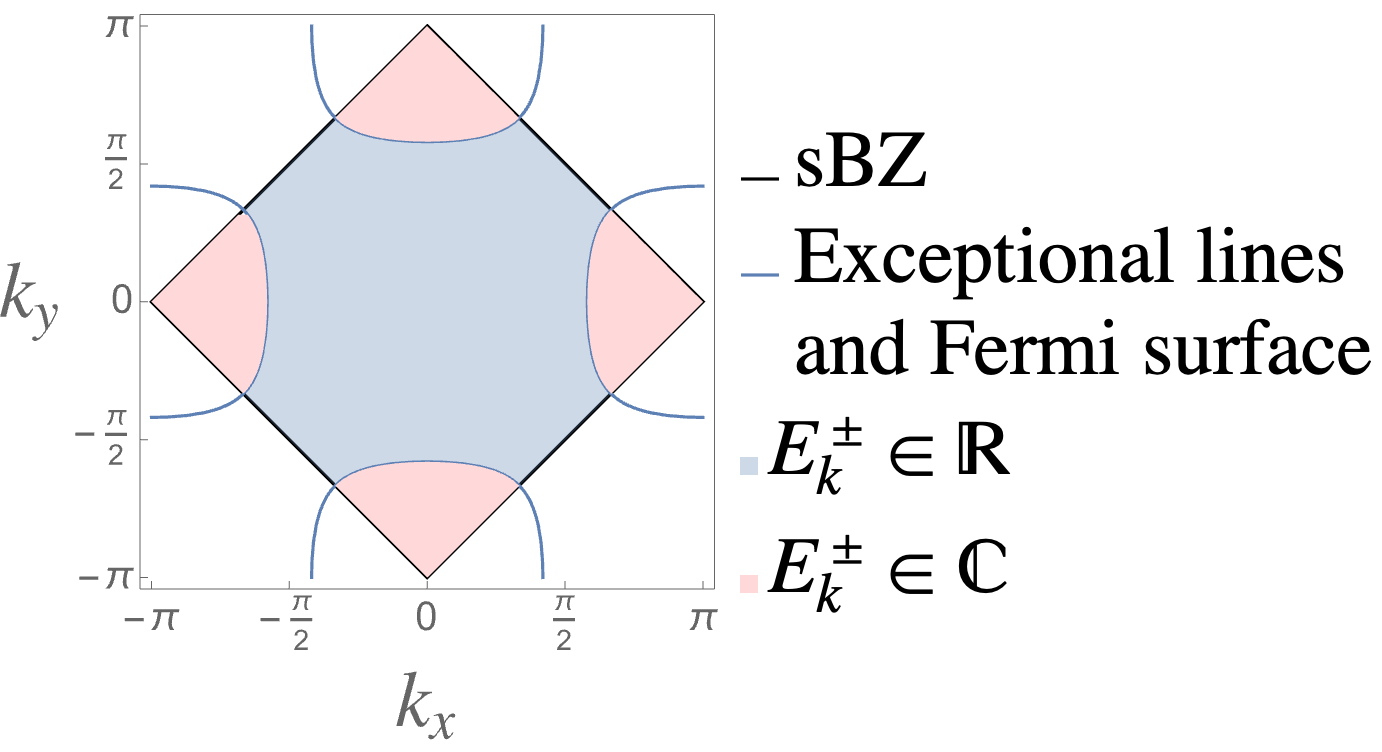}}
    \vfill
    \subfloat[]{\includegraphics[width=0.49\linewidth]{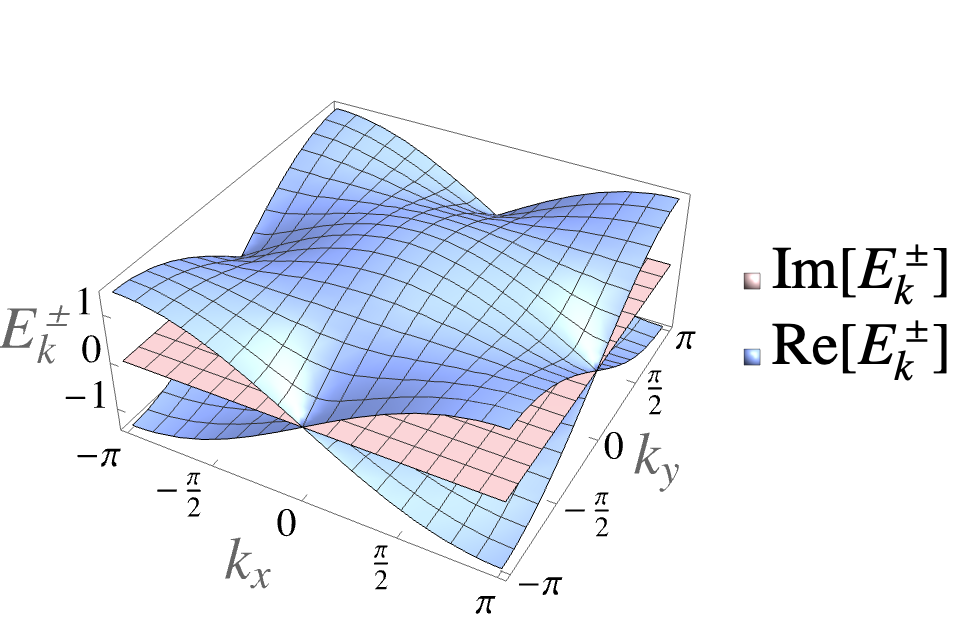}}
    \hspace{0.02cm}
    \subfloat[]{\includegraphics[width=0.49\linewidth]{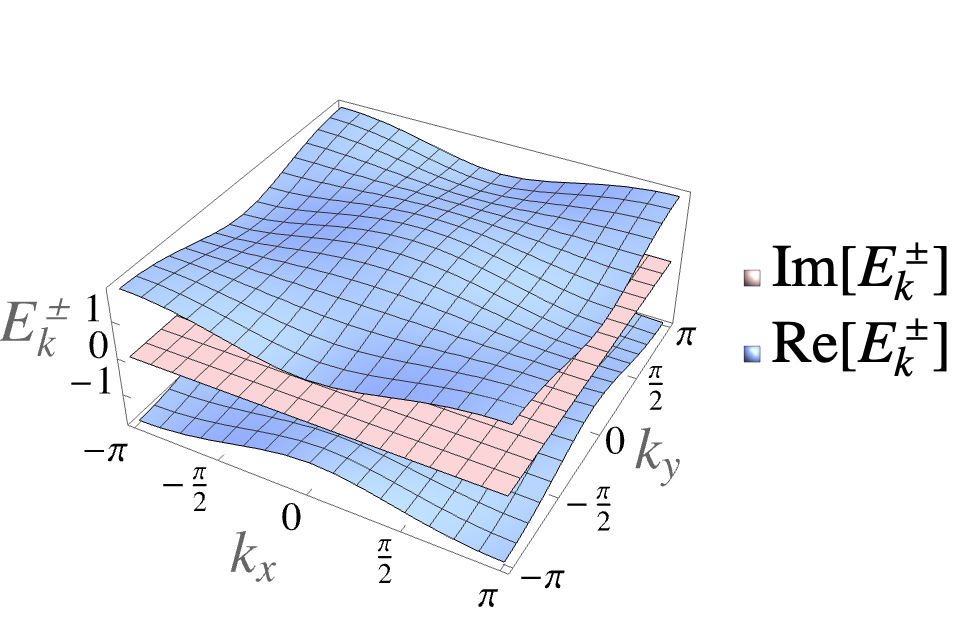}}
 	\caption{Calculated energy spectrum of the interacting two-dimensional HN flux model for $\mu=0$ of the AF metal at $W=0.5t$ (a), AF semimetal at $W=t$ (c) and AF insulator at $W=1.5t$ (d). Panel (b) shows the Fermi surface and exceptional lines for $W=0.5t$.} 
 	\label{Fig:Spectrum}
\end{figure}

In what follows, we consider the most interesting case of half-filling, $\mu=0$, and study the possible formation of the long-range $(\pi,\pi)$ antiferromagnetic order, mediated by the Hubbard on-site repulsion. Due to the sublattice structure of the model, the wave vector $\vec{Q}_{\text{AF}}=(\pi,\pi)$ is a reciprocal one, i.e. $\varepsilon_{\vec{k}+\vec{Q}_{\text{AF}}}=\varepsilon_{\vec{k}}$. After a Hubbard-Stratonovich transformation, Eq.~\eqref{H_NH} is written as
\begin{eqnarray}
	\label{H_MF}
	H_{\text{MF}}&=&H_t-\mu\sum_{\boldsymbol{r},\sigma}f^{\dagger}_{\boldsymbol{r},\sigma}f_{\boldsymbol{r},\sigma}+\frac{U}{2}\sum_{\boldsymbol{r}}m^2_{\boldsymbol{r}}\nonumber\\&-&U\sum_{\boldsymbol{r}}m_{\boldsymbol{r}}\left(f^{\dagger}_{\boldsymbol{r},\uparrow}f_{\boldsymbol{r},\uparrow}-f^{\dagger}_{\boldsymbol{r},\downarrow}f_{\boldsymbol{r},\downarrow}\right),
\end{eqnarray}
where $m_{\boldsymbol{r}}$ is the auxiliary field. At a mean-field level, we define
\begin{eqnarray}
\label{AFM}
	m_{\boldsymbol{r}}=\begin{cases}
	m, \quad \boldsymbol{r}\in A\\
	-m, \quad \boldsymbol{r}\in B
	\end{cases}
\end{eqnarray}
to describe the antiferromagnetic sublattice magnetization. Upon performing a Fourier transform, the mean-field Hamiltonian \eqref{H_MF} is diagonalized to reveal the energy spectrum,
\begin{eqnarray}
\label{Spectrum}
	E_{\vec{k}}^{\pm}=-\mu\pm\sqrt{\varepsilon^2_{\vec{k}}+W^2}.
\end{eqnarray}
where $W=U|m|$ and $\varepsilon_{\vec{k}}$ is as defined above in Eq.~\eqref{Eq:epsilon}. 

For $W<t$ the system is an AF metal with complex and real-valued energies depending on the values of the momentum, see Fig.~\ref{Fig:Spectrum}-(a,b). For $\mu=0$ the Fermi surface coincides with exceptional lines that separate real ($\mathcal{PT}$ symmetry is unbroken) and complex ($\mathcal{PT}$ symmetry is broken) valued energies. For $W \ge t$, the system undergoes a transition from an AF semimetal at $W=t$ (with Dirac points located at $(\pm \pi,0)$ and $(0,\pm \pi)$) to an AF insulator for $W>t$ with only real-valued energies Fig.~\ref{Fig:Spectrum}-(c,d), where the $\mathcal{PT}$ symmetry is restored in the entire BZ.

\begin{figure}[h!]
    \includegraphics[width=0.45\linewidth]{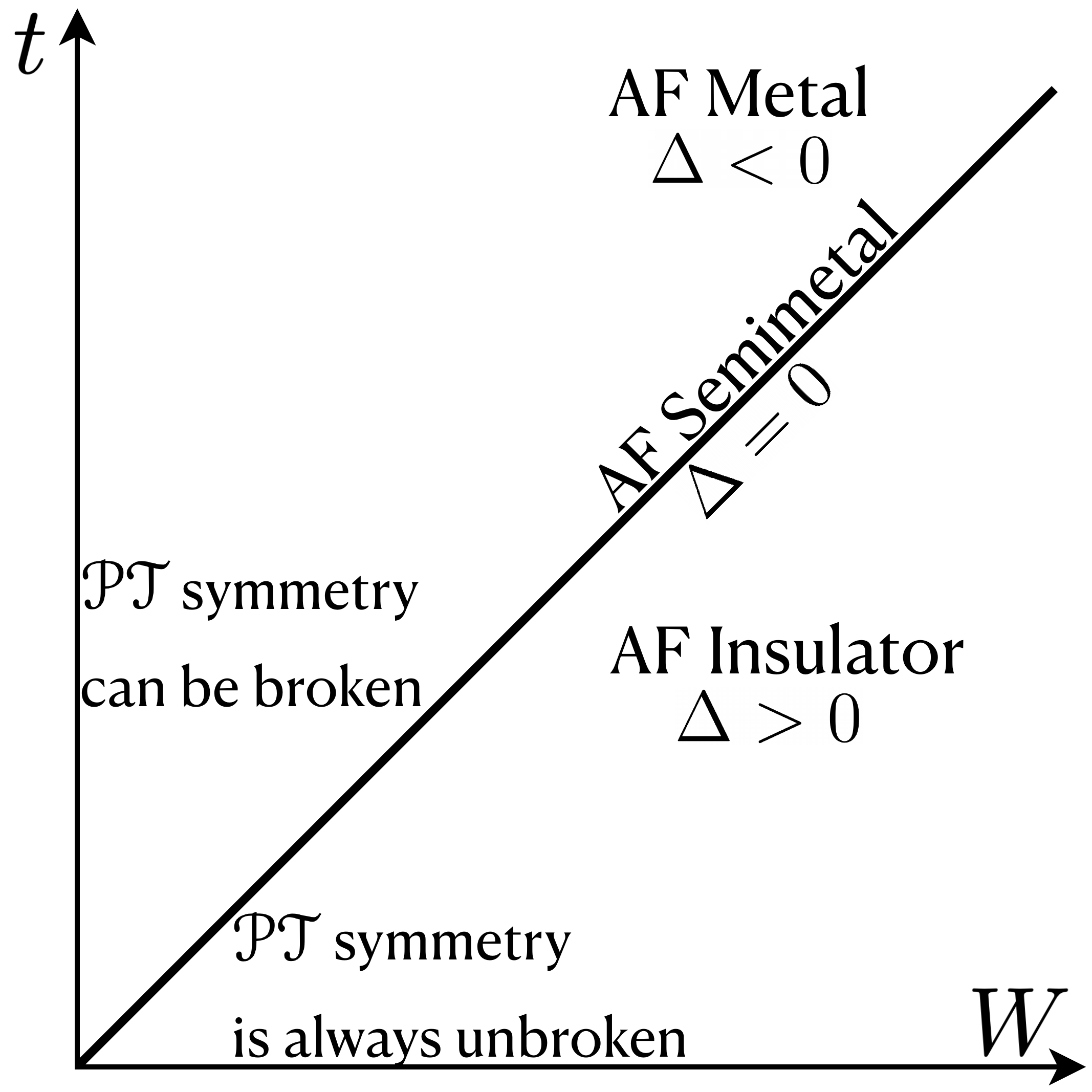}
 	\caption{Schematic phase diagram in terms of the gap order parameter $\Delta=W^2/t^2-1$ distinguishing the AF metal ($\Delta<0$), AF semimetal ($\Delta=0$) and AF insulator ($\Delta>0$).} 
 	\label{Fig:Phase_diagram}
\end{figure}

It is interesting that the change of variables, $k_x=k'_x+k'_y$ and $k_y=k'_x-k'_y$, transforms the energy dispersion (\ref{Spectrum}) to,
\begin{eqnarray}
\label{Spectrum-1}
	E_{\vec{k}'}^{\pm}=-\mu\pm t\sqrt{\cos^2k'_x+\cos^2k'_y+\Delta},
\end{eqnarray}
where $\Delta=W^2/t^2-1$ is the gap order parameter. A schematic phase diagram for the metal-insulator phase transition is shown in Fig.~\ref{Fig:Phase_diagram}. For the metallic phase $\Delta<0$ ($W<t$) and the $\mathcal{PT}$ symmetry can be broken for some values of momenta $(k'_x, k'_y)$. For the insulating phase $\Delta>0$ ($W>t$), with the $\mathcal{PT}$ symmetry remaining unbroken for all values of the momenta $(k'_x, k'_y)$. Finally, at the critical value $\Delta=0$ ($W=t$), we have a semimetal phase. The point for which $\Delta=0$ ($W=t$) Eq.~\eqref{Spectrum-1} coincides with the spectrum of the $\pi$-flux phase for the quantum antiferromagnet in a square lattice \cite{Affleck-Marston_PhysRevB.37.3774,Lieb_PhysRevLett.73.2158}. For $W=t$ the energy spectrum $E_{\vec{k}}^{\pm}$ near the four points $(\pm\pi,0)$, $(0,\pm\pi)$ in the corners of the sBZ zone is linear $E_{\vec{p}}^{\pm}=-\mu\pm v_F|\vec{p}|$, where vector $\vec{p}=(p_x,p_y)$ is counted from these four points, and the Fermi velocity is equal to $v_F=t/\sqrt{2}$.

\subsection{Topological triviality of the mean-field Hamiltonian}

In this section we briefly discuss the topological properties of our model. To do this we use the matrix representation of the mean-field Hamiltonian in Fourier space
\begin{eqnarray}
\label{Maxtix_H_MF}
	\hat{H}_{\text{MF}}(\boldsymbol{k})&=\begin{bmatrix}
	-\mu+\sigma W & -t\cos{k_y}\\
	-t\cos{k_x} & -\mu-\sigma W
	\end{bmatrix}
\end{eqnarray}
For such a matrix we can introduce a set of right and left eigenvectors
\begin{subequations}
    \begin{empheq}[]{align} 
       &\hat{H}_{\text{MF}}(\boldsymbol{k})\ket{R^{\pm}_{\vec{k},\sigma}}=E_{\vec{k}}^{\pm}\ket{R^{\pm}_{\vec{k},\sigma}},\\
       &\hat{H}^{\dagger}_{\text{MF}}(\boldsymbol{k})\ket{L^{\pm}_{\vec{k},\sigma}}=(E_{\vec{k}}^{\pm})^*\ket{L^{\pm}_{\vec{k},\sigma}}
    \end{empheq}
\end{subequations}
where $E_{\vec{k}}^{\pm}$ is given by the Eq.~\eqref{Spectrum}. Expressions for the left and right eigenvectors are presented in Appendix B.

Typically, non-trivial topological behavior is associated with a non-zero value of the Chern number. For the Hermitian model, we can define Chern number as \cite{Esaki, Shen}
\begin{eqnarray}
\label{Chern}
	\mathcal{C}^{\pm}=\frac{i}{2\pi}\int_{\text{sBZ}}\varepsilon^{ij}\bra{\partial_{k_i}L^{\pm}_{\vec{k},\sigma}}\ket{\partial_{k_j} R^{\pm}_{\vec{k},\sigma}}d^2k,
\end{eqnarray}
Using the expressions for left and right eigenvectors that we calculated in Appendix B, we can write
\begin{eqnarray}
	\hspace{-0.5cm}\mathcal{C}^{\pm}&=\pm&\frac{iWt^2}{8\pi}\int_{\text{sBZ}}dk_xdk_y\frac{\sin k_x\sin k_y}{(\varepsilon_{\vec{k}}^2+W^2)^{3/2}}=0.
\end{eqnarray}
The integral is zero because the integrand is odd function with respect to $k_x$ and $k_y$, and the sBZ is symmetric. This indicates trivial topological properties. However, we should not forget that in non-Hermitian models, phase transitions can occur without changing the Chern number. For example, a phase transition can occur when we pass through an exceptional point, or an exceptional line as in our case.

In addition to the Chern number, we can define for non-Hermitian model the winding number \cite{Leykam, Shen, Kawabata_2019_Aug}
\begin{equation}
    \nu^{\pm}=\oint\limits_{S}\frac{d\vec{k}}{2\pi i}\cdot\nabla_{\vec{k}}\log\det[H_{\text{MF}}(\vec{k})-E_{\vec{k}_{\text{EP}}}^{\pm}],
\end{equation}
where $S$ is a loop in momentum space that encircles one of the exceptional points ($\varepsilon^2_{\vec{k}_{\text{EP}}}=-W^2$). In our case, $\nu^{\pm}=0$, which again indicates the absence of topological phases.

%Although the spectrum is gapless for the case $W=t$, there is another way to generate a gap. Similarly to graphene \cite{Khveshchenko_PhysRevLett.87.246802,Gorbar_PhysRevB.66.045108}, we can assume that the two-dimensional system is actually embedded in a three-dimensional space and take into account the long-range Coulomb interaction with a potential $V(\vec{r}-\vec{r}')=e^2/(\varepsilon|\vec{r}-\vec{r}'|)$. In this case, a gap is dynamically generated.  This follows by truncating the Schwinger-Dyson equation with the Coulomb interaction as the vertex, which leads to the self-consistent equation, 
%\begin{eqnarray}
%\label{Gap_Eq}
%	\Sigma(\vec{q})=\frac{\alpha}{4\pi}\int d^2k\frac{\Sigma(\vec{k})}{\sqrt{\vec{k}^2+\Sigma^2(\vec{k})}|\vec{k}+\vec{q}|},
%\end{eqnarray}
%where $\alpha=e^2/(\varepsilon v_F)$ is the "fine-structure constant" of the system and $\Sigma(\vec{k})$ is the self-energy. After some simplifications, we can find the following approximate solution to Eq.~\eqref{Gap_Eq}
%\begin{eqnarray}
%	\Sigma(0)=c\Lambda\exp[-\frac{\text{const}}{\sqrt{2\alpha-1}}],
%\end{eqnarray}
%where $c$ is a constant and $\Lambda$ is an ultraviolet cutoff. From this we can see that the constant $\alpha$ has a critical value $\alpha_c=1/2$ above which the energy spectrum acquires a gap. 

\section{Dynamic spin susceptibility}
\label{Dynamic_spin_susceptibility}

To investigate the stability of the long-range $(\pi,\pi)$ AF order and the properties of the AF state we continue further by analyzing the longitudinal and transverse (spin wave) fluctuations of the magnetically ordered state. Usually, models with non-Hermitian excitation require a modified response function \cite{Pan, Sticlet, Geier, Ilya, Sim}. However, in our case, although the model itself is non-Hermitian, the excitation is Hermitian. Therefore, we can use the usual definition for bare dynamical spin susceptibility for the transverse, $+-$, and the  longitudinal, $zz$, components
\vspace{-0.02cm}
\begin{eqnarray}
	\label{bare_susceptibility}
    \chi^{lm}_0(\boldsymbol{q},\boldsymbol{q}',\Omega)=\int dt e^{i\Omega t}\langle T_{t} S^l_{\boldsymbol{q}}(t)S^m_{-\boldsymbol{q}'}(0)\rangle,
\end{eqnarray}
where $lm=+-,zz$, with $S^\pm=S^x\pm iS^y$ corresponding to the usual spin ladder operators, $\Omega$ is a frequency and $T_t$ is a time ordering operator.

Within the random phase approximation (RPA) framework, the total susceptibility is expressed as a sum of bubble diagrams, where each bubble represents the susceptibility of a non-interacting system calculated with respect to the mean-field ground state. 

Normally RPA summation is a $2\times2$ matrix problem since the antiferromagnetic ordering at $\vec{q}=(\pi,\pi)$ doubles the unit cell, but our non-Hermitian model consists of two sublattices from the very beginning. As a consequence there are no extra umklapp terms appearing in the AF state. Therefore, within RPA both components of the susceptibility can be expressed as, 
\begin{eqnarray}
	\label{RPA}
	\chi^{lm}_{\text{RPA}}(\boldsymbol{q},\boldsymbol{q},\Omega)=\frac{\chi^{lm}_{0}(\boldsymbol{q},\boldsymbol{q},\Omega)}{1-U\chi^{lm}_{0}(\boldsymbol{q},\boldsymbol{q},\Omega)},
\end{eqnarray}
with the bare components being found as
\begin{widetext}
\vspace{0.2cm}
\begin{eqnarray}
\label{chi_0+-}
	\chi^{+-}_0(\boldsymbol{q},\boldsymbol{q},\Omega)&=&\frac{1}{2V}\sum_{\boldsymbol{k},c}{}'\left(1-\frac{W^2-t^2/2(\cos k_x\cos(k_y+q_y)+\cos(k_x+q_x)\cos k_y)}{\sqrt{\varepsilon^2_{\boldsymbol{k}}+W^2}\sqrt{\varepsilon^2_{\boldsymbol{k}+\boldsymbol{q}}+W^2}}\right)\frac{f(E^{c}_{\boldsymbol{k}+\boldsymbol{q}})-f(E^{c}_{\boldsymbol{k}})}{\Omega+i0^{+}-E^{c}_{\boldsymbol{k}+\boldsymbol{q}}+E^{c}_{\boldsymbol{k}}}\nonumber\\
	&+&\frac{1}{2V}\sum_{\boldsymbol{k},c\neq c'}{}^{'}\left(1+\frac{W^2-t^2/2(\cos k_x\cos(k_y+q_y)+\cos(k_x+q_x)\cos k_y)}{\sqrt{\varepsilon^2_{\boldsymbol{k}}+W^2}\sqrt{\varepsilon^2_{\boldsymbol{k}+\boldsymbol{q}}+W^2}}\right)\frac{f(E^{c'}_{\boldsymbol{k}+\boldsymbol{q}})-f(E^{c}_{\boldsymbol{k}})}{\Omega+i0^{+}-E^{c'}_{\boldsymbol{k}+\boldsymbol{q}}+E^{c}_{\boldsymbol{k}}},
\end{eqnarray}
and
\begin{eqnarray}
\label{chi_0zz}
	\chi^{zz}_0(\boldsymbol{q},\boldsymbol{q},\Omega)&=&\frac{1}{V}\sum_{\boldsymbol{k},c}{}'\left(1+\frac{W^2+t^2/2(\cos k_x\cos(k_y+q_y)+\cos(k_x+q_x)\cos k_y)}{\sqrt{\varepsilon^2_{\boldsymbol{k}}+W^2}\sqrt{\varepsilon^2_{\boldsymbol{k}+\boldsymbol{q}}+W^2}}\right)\frac{f(E^{c}_{\boldsymbol{k}+\boldsymbol{q}})-f(E^{c}_{\boldsymbol{k}})}{\Omega+i0^{+}-E^{c}_{\boldsymbol{k}+\boldsymbol{q}}+E^{c}_{\boldsymbol{k}}}\nonumber\\
	&+&\frac{1}{V}\sum_{\boldsymbol{k},c\neq c'}{}^{'}\left(1-\frac{W^2+t^2/2(\cos k_x\cos(k_y+q_y)+\cos(k_x+q_x)\cos k_y)}{\sqrt{\varepsilon^2_{\boldsymbol{k}}+W^2}\sqrt{\varepsilon^2_{\boldsymbol{k}+\boldsymbol{q}}+W^2}}\right)\frac{f(E^{c'}_{\boldsymbol{k}+\boldsymbol{q}})-f(E^{c}_{\boldsymbol{k}})}{\Omega+i0^{+}-E^{c'}_{\boldsymbol{k}+\boldsymbol{q}}+E^{c}_{\boldsymbol{k}}},
\end{eqnarray}
\vspace{0.2cm}
\end{widetext}
where $c=\pm$ are the band indices, $f(\varepsilon)$ is the Fermi-Dirac distribution and $V$ is the volume. The prime refers to the sum over the (reduced) sBZ. For details, see Appendix A.

Like in the Hermitian single-band Hubbard model \cite{Chubukov92}, the spin-rotational invariance is broken, $\chi^{zz}_{\text{RPA}}\neq 2\chi^{+-}_{\text{RPA}}$. While $\chi^{zz}_{\text{RPA}}$ at the ordering wavevector is gapped up to twice the AF gap energy, $\chi^{+-}_{\text{RPA}}$ determines the spectrum of the spin waves. In particular, for $\Omega=0$ and $\boldsymbol{q}=\boldsymbol{Q}_{\text{AF}}=(\pi,\pi)$, the RPA expression for transverse component has a pole
\begin{eqnarray}
\label{Pole}
	1-U\chi^{+-}_{0}(\boldsymbol{Q}_{\text{AF}},\boldsymbol{Q}_{\text{AF}},0)=0,
\end{eqnarray}
which corresponds to the gapless Goldstone mode and is equivalent to the mean-field equation that determines the mean-field AF order,
\begin{eqnarray}
\label{Self_consistent_Eq}
	\frac{1}{U}=-\frac{1}{V}\sum_{\boldsymbol{k}}{}^{'}\frac{1}{\sqrt{\varepsilon^2_{\boldsymbol{k}}+W^2}}\left[f(E^{+}_{\boldsymbol{k}})-f(E^{-}_{\boldsymbol{k}})\right].
\end{eqnarray}
In Appendix B we show that the same result can be obtained using the partition function in bi-orthogonal basis.

\begin{figure}
    \subfloat[]
    {\includegraphics[width=0.5\linewidth]{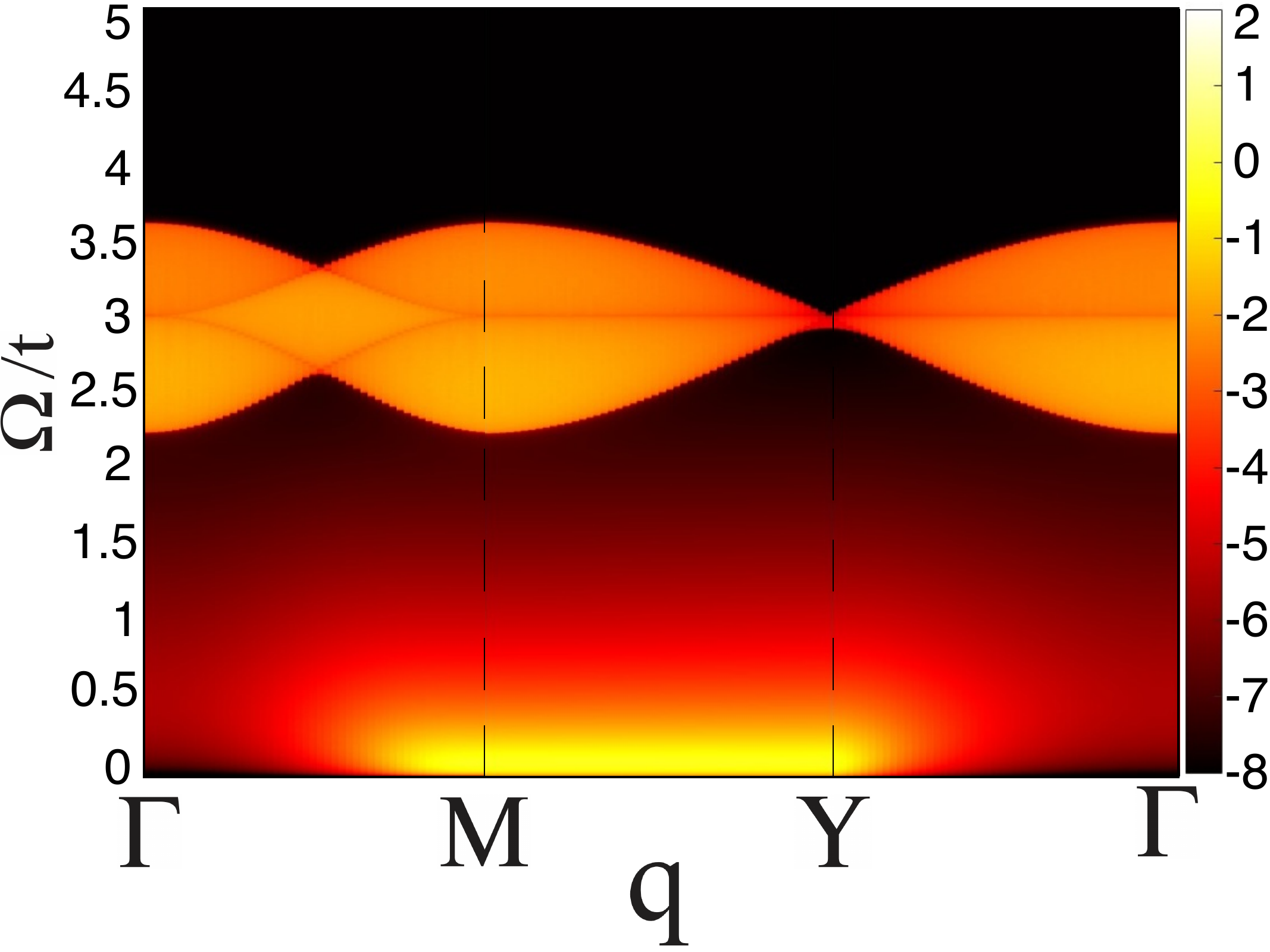}}
    \subfloat[]
    {\includegraphics[width=0.5\linewidth]{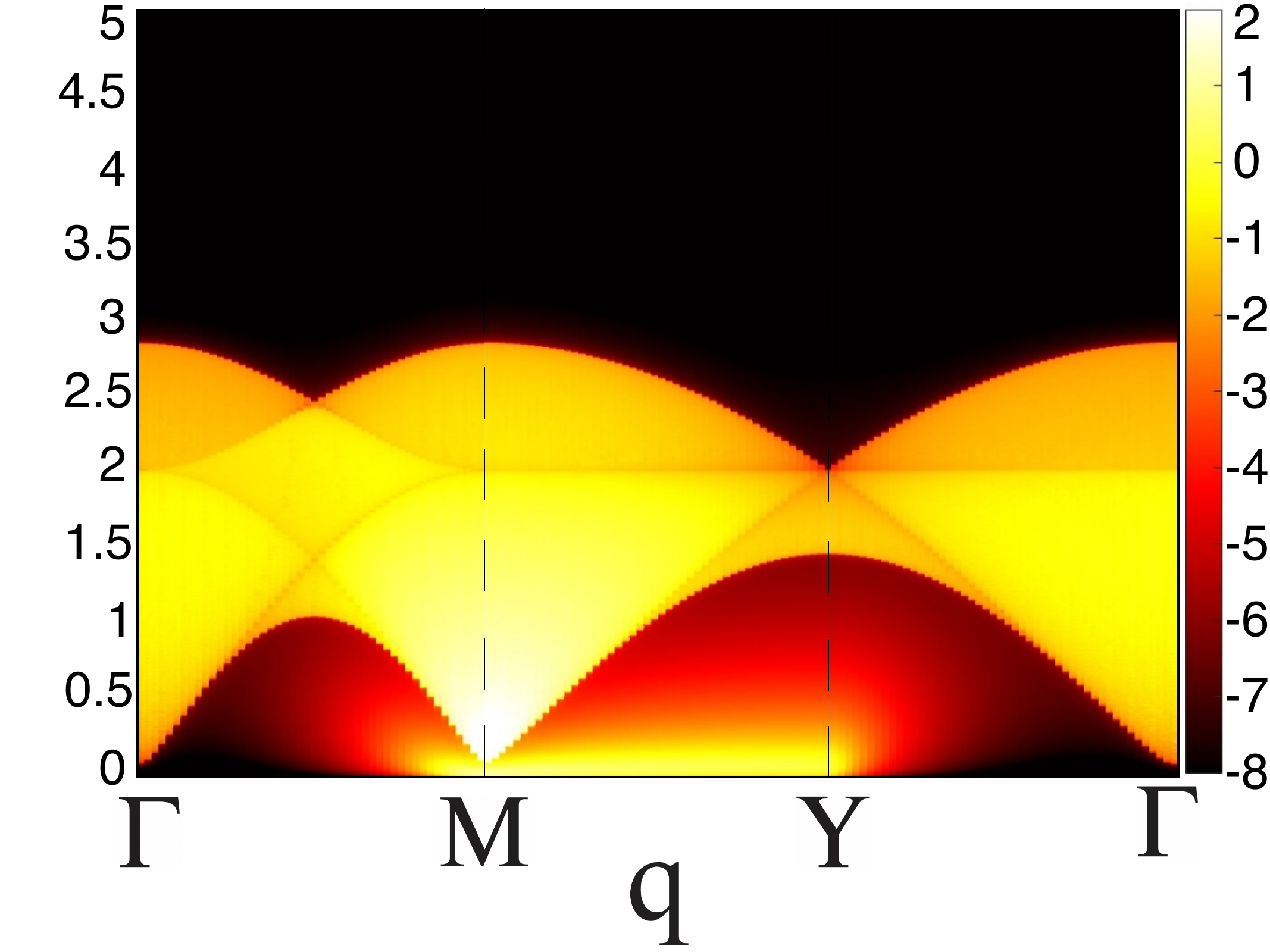}}
 	\caption{Calculated transverse component of the RPA spin excitation spectrum along the symmetry route $(0,0)\to(\pi,\pi)\to(0,\pi)\to(0,0)$ of the BZ for (a) AF insulator $W=1.5t$, $\mu=0$, $U=1.47028t$ and (b) AF semimetal $W=1.001t$, $\mu=0$, $U=0.79545t$. Here, we calculate $U$ self-consistently for a given $W$. Intensity of $\text{Im}\chi^{+-}_{\text{RPA}}(\boldsymbol{q},\boldsymbol{q},\Omega)$ shown on the log scale.} 
 	\label{Fig:Spin-wave_dispersion}
\end{figure}

The solution of Eq.~\eqref{Self_consistent_Eq} determines the parameter $W$ self-consistently for a given $U$ and as a function of the temperature. Furthermore, note that Eq.~\eqref{Self_consistent_Eq} is only well-defined for the real-valued energy spectrum for $W>t$, that is for an AF insulator and AF semimetal. 
%For the AF metal, $W<t$, and the eigenenergies become complex-valued at some points of the BZ. Then the solution is only possible if an imaginary component of the interaction is assumed, i.e. if $U=\sqrt{(\mbox{Re}U)^2+(\delta U)^2}$  where $\delta U$ is a dynamically generated imaginary component of the interaction. This provides an interesting route to generate non-Hermitian magnons for the AF metal, which  we will study this elsewhere.

In what follows we proceed with $\mathcal{PT}$-symmetric case. In Fig.~\ref{Fig:Spin-wave_dispersion} we show the imaginary part of the RPA spin susceptibility along the high-symmetry points of the BZ for the AF insulator employing $W=1.5t$ (a) and AF semimetal $W=t$ (b). The situation resembles the Hermitian case and the collective spin excitations are $\mathcal{PT}$-symmetric. One immediately sees that the collective modes (spin waves) are undamped, gapless, and well-separated from the particle-hole continuum band, centered around $3t$. Furthermore, spin waves are nearly flat along M -- X as well as M -- Y direction, which signals about potential instability of the $(\pi,\pi)$ mean-field AF order against fluctuations. To see this analytically we expand the denominator of Eq.~\eqref{RPA} around $\boldsymbol{q}=\boldsymbol{Q}_{\text{AF}}$, $\Omega=0$ up to quadratic order
\begin{eqnarray}
    \label{Expansion}
	\chi^{+-}_{\text{RPA}}(\boldsymbol{q},\boldsymbol{q},\Omega)\cong-\frac{\chi^{+-}_{0}(\boldsymbol{q},\boldsymbol{q},\Omega)}{c_1\Omega^2+c_2(\delta q_x)^2(\delta q_y)^2},
\end{eqnarray}
where coefficients $c_1$ and $c_2$ are positive and defined as
\begin{eqnarray}
	c_1=\frac{1}{4V}\sum_{\boldsymbol{k}}{}^{'}\frac{1}{(\varepsilon^2_{\boldsymbol{k}}+W^2)^{3/2}},
\end{eqnarray}
and
\begin{widetext}
\begin{eqnarray}
	c_2&=&\frac{1}{128V}\sum_{\boldsymbol{k}}{}^{'}\frac{1}{(\varepsilon^2_{\vec{k}}+W^2)^{9/2}}\big[t^4\sin^2k_x\sin^2k_y(48W^4-64W^2\varepsilon^2_{\vec{k}}-7\varepsilon^4_{\vec{k}})\nonumber\\
    &-&6t^4(\cos^2k_x\sin^2k_y+\sin^2k_x\cos^2k_y)(6W^4+7W^2\varepsilon^2_{\vec{k}}+\varepsilon^4_{\vec{k}})-4\varepsilon^2_{\vec{k}}(4W^6+9W^4\varepsilon^2_{\vec{k}}+6W^2\varepsilon^4_{\vec{k}}+\varepsilon^6_{\vec{k}})\big],
\end{eqnarray}
\end{widetext}
and $\delta\vec{q}=\vec{q}-\vec{Q}_{\text{AF}}$.
Equation \eqref{Expansion} has poles at $\Omega(\vec{q})=\pm i\sqrt{c_2/c_1}\delta q_x\delta q_y$, which are purely imaginary, in contrast with ordinary forms of antiferromagnetic excitations featuring a real spectrum linear in the magnitude of the wave vector.  On the other hand, an imaginary dispersion relation of the form $\Omega(q)=\pm c_s|\vec{q}|-iDq^2$ with $c_s,D>0$ describes AF spin waves including dissipation from heat diffusion \cite{Hohenberg}. In our case the real part vanishes and the negative sign corresponds to a purely diffusive mode of a $d$-wave type. However, the positive sign leads to the opposite of a damping, so both signs taken together can be interpreted as being associated to gain and loss, in a behavior inherent to non-Hermitian systems. Usually, diffusive modes of the ordinary type naturally occur for the paramagnetic state \cite{Hohenberg}. The uncovered gain and loss process is finite only in a small interval of $W$ values and quickly approaches zero as $W$ increases. Furthermore, the coefficients $c_1$ and $c_2$ are well defined only in the region of unbroken $\mathcal{PT}$ symmetry ($W>t$), in which case the energy spectrum \eqref{Spectrum} is real. Usually in this case we are not dealing with such non-unitary phenomena as gain and loss. However, for our model the region of unbroken $\mathcal{PT}$ symmetry corresponds to the imaginary dispersion relation for spin waves. This means that the gain and loss mechanisms in this case take place even in the region where $\mathcal{PT}$ symmetry is unbroken.

The spin waves remain nearly flat also in the case of the AF semimetal and do not interfere with the particle-hole continuum, which is also gapless in this case. We conclude that the long-range $(\pi,\pi)$ AF order is highly frustrated due to nearly flat spin wave excitations spectrum. Note that the potential instability of the system towards the stripe-type order $(\pi,0)$ (or $(0,\pi)$) could be seen from the properties of the tight-binding dispersion at $\vec{Q}_{\text{X}} =(\pi,0)$, $\varepsilon_{\vec{k}+\vec{Q}_{\text{X}}}=i\varepsilon_{\vec{k}}$, which can be regarded as non-Hermitian nesting.

\section{Conclusions}
To conclude, we have formulated the non-Hermitian flux model where each plaquette on a square lattice consists of a spinful, maximally asymmetric Hatano-Nelson model. As a consequence, the model is characterized by the clock- and anticlockwise asymmetric hoppings on a plaquette within the square lattice and resembles the flux phase discussed in the context of $U(1)$ spin liquid phases of AF $SU(N)$ Heisenberg models in the large $N$ limit \cite{Affleck-Marston_PhysRevB.37.3774}. At half-filling the non-interacting model describes a metal characterized by a Fermi surface with simultaneous exceptional lines separating the real- and complex-valued branches of the energy spectrum. After including an on-site Coulomb repulsion we analyzed the formation of a long-range AF phase with $(\pi,\pi)$ ordering wave vector at half-filling. We find that for the AF insulator the AF order restores the $\mathcal{PT}$ symmetry with a real-valued energy spectrum. At the same time, for the AF metal, $\mathcal{PT}$ symmetry can be broken for certain values of momentum.

The transverse spin excitations (spin waves) are gapless and nearly flat along the M -- Y and M -- X directions of the BZ indicating  potential frustration of $(\pi,\pi)$ AF order towards $(\pi,0)$ or $(0,\pi)$ states. This instability was demonstrated to be associated with the emergence of diffusive modes of a $d$-wave type. However, the diffusion is in this case does not only include a dissipative contribution (loss), but also an enhancement mode (gain), which is a consequence of the non-Hermiticity of the model. Moreover, this behavior occurs in the region of unbroken $\mathcal{PT}$ symmetry. Usually the non-standard behavior of non-Hermitian $\mathcal{PT}$ symmetric systems is always associated with the region where the $\mathcal{PT}$ symmetry is broken. Here we are dealing with a fundamentally new situation, where the phenomena of gain and loss occur even in the region of unbroken $\mathcal{PT}$ symmetry, as the evidenced by the behavior of the dynamic susceptibility. This indicates that even in the region of unbroken $\mathcal{PT}$ symmetry, one can expect phenomena related to the interaction of the system with the environment. 

From an experimental point of view, we can expect, for example, dissipative behavior within systems similar to ours even in a region of unbroken $\mathcal{PT}$ symmetry.

We believe that there are two fairly simple ways to implement our model. The first method consists in the experimental realization of a non-Hermitian Hamiltonian featuring a non-Hermitian skin eﬀect (NHSE) in a momentum lattice using ultracold atoms \cite{Meier2016,Meier2018,Lapp2019,Xie2019}. Namely, there is a possibility of creating a chain of coupled Aharonov-Bohm (AB) rings along a synthetic momentum lattice \cite{Liang} using Bose-Einstein condensate of $^{87}$Rb atoms \cite{Xie2018,Xie2020,Xie2021}.

The second method involves ready-made two-dimensional non-reciprocal topological electric circuits \cite{Stegmaier, Helbig, Liu} with slightly modified hopping according to the definition of the flux model.

\begin{acknowledgments}
	We acknowledge financial support by the Deutsche Forschungsgemeinschaft (DFG, German Research Foundation), through SFB 1143 project A5 and the W{\"u}rzburg-Dresden Cluster of Excellence on Complexity and Topology in Quantum Matter-ct.qmat (EXC 2147, Project Id No. 390858490). 
\end{acknowledgments}

\label{Conclusions_and_outlook}

\appendix
\section{Derivation of the dynamical spin susceptibility}
To calculate the bare dynamical spin susceptibility for the transverse and the longitudinal components, we first need to obtain the matrix elements of the Green's function for the mean-field Hamiltonian \eqref{Maxtix_H_MF}
\begin{eqnarray}
	\hat{G}(\boldsymbol{k},i\omega_n)&=&
(-i\omega_n\mathbbm{1}-\hat{H}_{\text{MF}}(\boldsymbol{k}))^{-1},
\end{eqnarray}
in order to obtain matrix elements of Green's function, 
\begin{subequations}
\label{Green}
    \begin{empheq}[]{align}
        &G_{a^{\dagger}_{\sigma}a_{\sigma}}(\boldsymbol{k},i\omega_n)=-\frac{u_{\vec{k},\sigma}}{i\omega_n-E^+_{\boldsymbol{k}}}-\frac{v_{\vec{k},\sigma}}{i\omega_n-E^-_{\boldsymbol{k}}},\\
        &G_{b^{\dagger}_{\sigma}b_{\sigma}}(\boldsymbol{k},i\omega_n)=-\frac{v_{\vec{k},\sigma}}{i\omega_n-E^+_{\boldsymbol{k}}}-\frac{u_{\vec{k},\sigma}}{i\omega_n-E^-_{\boldsymbol{k}}},\\
        &G_{a^{\dagger}_{\sigma}b_{\sigma}}(\boldsymbol{k},i\omega_n)=w_{\vec{k}}\left(\frac{1}{i\omega_n-E^+_{\boldsymbol{k}}}-\frac{1}{i\omega_n-E^-_{\boldsymbol{k}}}\right),\\
        &G_{b^{\dagger}_{\sigma}a_{\sigma}}(\boldsymbol{k},i\omega_n)=x_{\vec{k}}\left(\frac{1}{i\omega_n-E^+_{\boldsymbol{k}}}-\frac{1}{i\omega_n-E^-_{\boldsymbol{k}}}\right),
    \end{empheq}
\end{subequations}
where
\begin{eqnarray}
	u_{\vec{k},\sigma}&=&\frac{1}{2}\left(1+\frac{\sigma W}{\sqrt{\varepsilon^2_{\vec{k}}+W^2}}\right), \: w_{\vec{k}}=\frac{t\cos k_y}{2\sqrt{\varepsilon^2_{\vec{k}}+W^2}},\hspace{0.5cm}\label{u_w}\\
    v_{\vec{k},\sigma}&=&\frac{1}{2}\left(1-\frac{\sigma W}{\sqrt{\varepsilon^2_{\vec{k}}+W^2}}\right), \: x_{\vec{k}}=\frac{t\cos k_x}{2\sqrt{\varepsilon^2_{\vec{k}}+W^2}}.\label{v_x}
\end{eqnarray}

We can now proceed directly to the definition of bare dynamical spin susceptibility \eqref{bare_susceptibility} and write  down the relation between the fermionic and spin operators,
\begin{subequations}
\label{Spins}
    \begin{empheq}[]{align} 
        &S^+_{\boldsymbol{q}}=\frac{1}{\sqrt{V}}\sum_{\boldsymbol{k}}{}^{'}\left(a^{\dagger}_{\boldsymbol{k}+\boldsymbol{q},\uparrow} a_{\boldsymbol{k},\downarrow}+b^{\dagger}_{\boldsymbol{k}+\boldsymbol{q},\uparrow} b_{\boldsymbol{k},\downarrow}\right),\\
        &S^-_{\boldsymbol{q}}=\frac{1}{\sqrt{V}}\sum_{\boldsymbol{k}}{}^{'}\left(a^{\dagger}_{\boldsymbol{k}+\boldsymbol{q},\downarrow} a_{\boldsymbol{k},\uparrow}+b^{\dagger}_{\boldsymbol{k}+\boldsymbol{q},\downarrow} b_{\boldsymbol{k},\uparrow}\right),\\
        &S^z_{\boldsymbol{q}}=\frac{1}{\sqrt{V}}\sum_{\boldsymbol{k}}{}^{'}\Big(a^{\dagger}_{\boldsymbol{k}+\boldsymbol{q},\uparrow} a_{\boldsymbol{k},\uparrow}-a^{\dagger}_{\boldsymbol{k}+\boldsymbol{q},\downarrow} a_{\boldsymbol{k},\downarrow}\nonumber\\
        &\hspace{0.5cm}+b^{\dagger}_{\boldsymbol{k}+\boldsymbol{q},\uparrow} b_{\boldsymbol{k},\uparrow}-b^{\dagger}_{\boldsymbol{k}+\boldsymbol{q},\downarrow} b_{\boldsymbol{k},\downarrow}\Big).
    \end{empheq}
\end{subequations}
By substituting Eqs.~\eqref{Spins} into the dynamical spin susceptibilities and using Vick's theorem, we can represent Eq.~\eqref{bare_susceptibility} in terms of the corresponding matrix elements of the Green's function, 
\begin{widetext}
\begin{eqnarray}
\label{chi_0+-G}
	\chi^{+-}_0(\boldsymbol{q},\boldsymbol{q},i\Omega_n)&=&
    \frac{1}{\beta V}\sum_{\boldsymbol{k},m}{}'\bigg[G_{a^{\dagger}_{\downarrow}a_{\downarrow}}(\boldsymbol{k},i\omega_m)G_{a^{\dagger}_{\uparrow}a_{\uparrow}}(\boldsymbol{k}+\boldsymbol{q},i\omega_m+i\Omega_n)+G_{b^{\dagger}_{\downarrow}b_{\downarrow}}(\boldsymbol{k},i\omega_m)G_{b^{\dagger}_{\uparrow}b_{\uparrow}}(\boldsymbol{k}+\boldsymbol{q},i\omega_m+i\Omega_n)\nonumber\\
	&+&G_{a^{\dagger}_{\downarrow}b_{\downarrow}}(\boldsymbol{k},i\omega_m)G_{b^{\dagger}_{\uparrow}a_{\uparrow}}(\boldsymbol{k}+\boldsymbol{q},i\omega_m+i\Omega_n)+G_{b^{\dagger}_{\downarrow}a_{\downarrow}}(\boldsymbol{k},i\omega_m)G_{a^{\dagger}_{\uparrow}b_{\uparrow}}(\boldsymbol{k}+\boldsymbol{q},i\omega_m+i\Omega_n)\bigg]
\end{eqnarray}
and
\begin{eqnarray}
\label{chi_0zzG}
    \chi^{zz}_0(\boldsymbol{q},\boldsymbol{q},i\Omega_n)&=&\frac{1}{\beta V}\sum_{\boldsymbol{k},m,\sigma}{}'\bigg[G_{a^{\dagger}_{\sigma}a_{\sigma}}(\boldsymbol{k},i\omega_m)G_{a^{\dagger}_{\sigma}a_{\sigma}}(\boldsymbol{k}+\boldsymbol{q},i\omega_m+i\Omega_n)+G_{b^{\dagger}_{\sigma}b_{\sigma}}(\boldsymbol{k},i\omega_m)G_{b^{\dagger}_{\sigma}b_{\sigma}}(\boldsymbol{k}+\boldsymbol{q},i\omega_m+i\Omega_n)\nonumber\\
	&+&G_{a^{\dagger}_{\sigma}b_{\sigma}}(\boldsymbol{k},i\omega_m)G_{b^{\dagger}_{\sigma}a_{\sigma}}(\boldsymbol{k}+\boldsymbol{q},i\omega_m+i\Omega_n)+G_{b^{\dagger}_{\sigma}a_{\sigma}}(\boldsymbol{k},i\omega_m)G_{a^{\dagger}_{\sigma}b_{\sigma}}(\boldsymbol{k}+\boldsymbol{q},i\omega_m+i\Omega_n)\bigg],
\end{eqnarray}
\end{widetext}
where $\Omega_n=2n\pi/\beta$ and $\omega_n=(2n+1)\pi/\beta$, $n\in\mathbb{Z}$ are bosonic and fermionic Matsubara frequencies, respectively and $\beta=1/T$ is the inverse of temperature. As a result, we have a set of bubble diagrams. Let's make detailed calculations of one of them. The rest of the diagrams can be calculated in a similar way.

\begin{figure}[t!]
\includegraphics[width=0.8\linewidth]{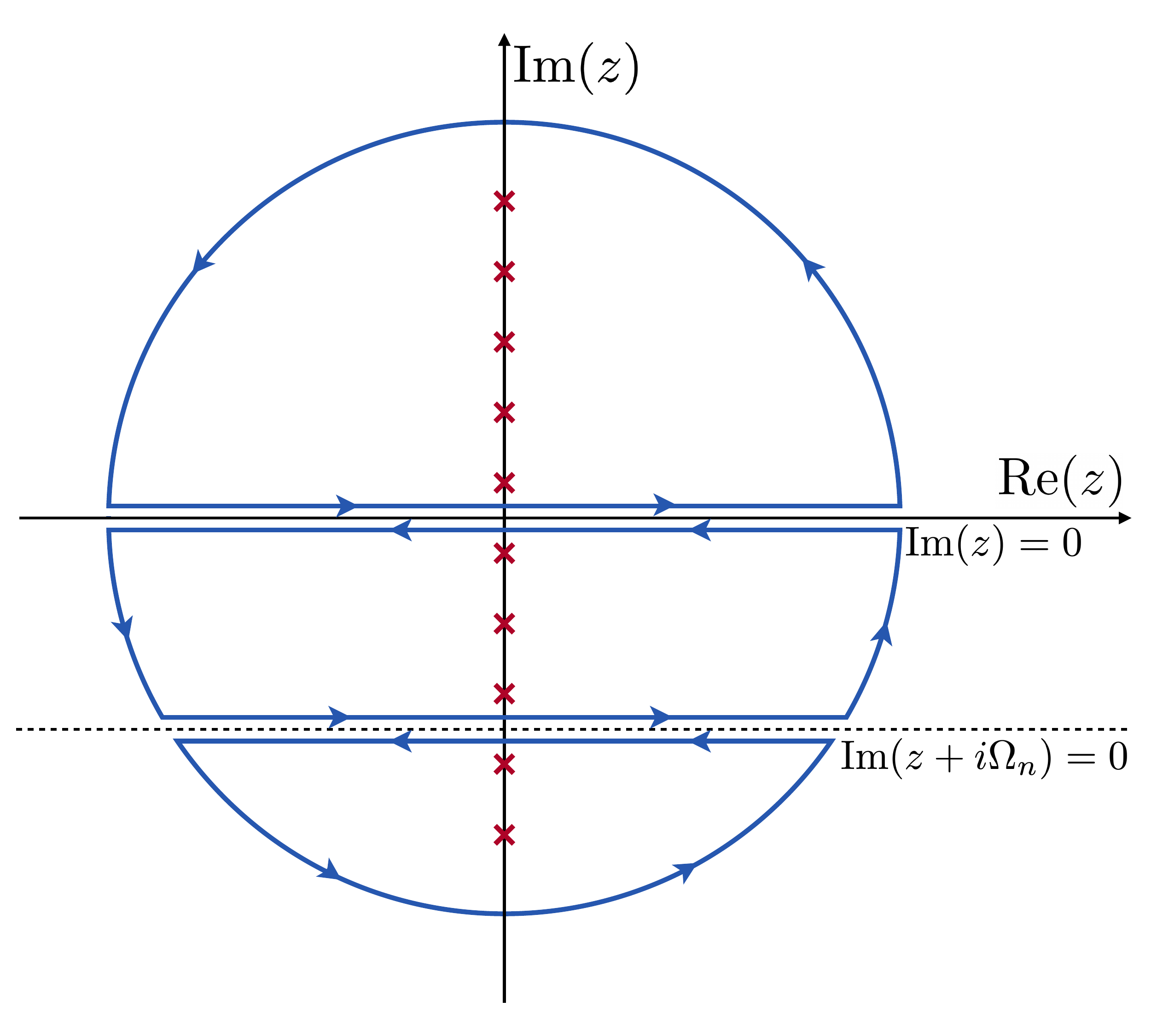}
 	\caption{The integration contour $C$ is represented by a blue line. The dashed lines indicate two branch cuts in the complex plane: $\text{Im}(z)=0$ and $\text{Im}(z+i\Omega_n)=0$. The red crosses correspond to the poles of Fermi-Dirac distribution function.} 
 	\label{Fig:Contour}
\end{figure}

First we can replace the Matsubara frequency summation by a contour integral
\begin{eqnarray}
	\frac{1}{\beta}\sum_{m}G_{a^{\dagger}_{\downarrow}a_{\downarrow}}(\boldsymbol{k},i\omega_m)G_{a^{\dagger}_{\uparrow}a_{\uparrow}}(\boldsymbol{k}+\boldsymbol{q},i\omega_m+i\Omega_n)\nonumber\\
    \hspace{-0.5cm}=-\oint\limits_{C}\frac{dz}{2\pi i}f(z)G_{a^{\dagger}_{\downarrow}a_{\downarrow}}(\boldsymbol{k},z)G_{a^{\dagger}_{\uparrow}a_{\uparrow}}(\boldsymbol{k}+\boldsymbol{q},z+i\Omega_n),
\end{eqnarray}
where $f(z)$ is the Fermi-Dirac distribution function. The contour is shown in Fig.~\ref{Fig:Contour}. In his contour integral, we need to avoid two horizontal branch cuts defined by $\text{Im}(z)=0$ and $\text{Im}(z+i\Omega_n)=0$. To calculate the integral, we note that the semicircular arcs do not contribute to the integral, and therefore only the contribution from the horizontal lines remains. Now we can parameterize the upper branch cut as $z=\epsilon+i0^+$, $z=\epsilon-i0^+$ and the lower cut as $z=\epsilon-i\Omega_n+i0^+$, $z=\epsilon-i\Omega_n-i0^+$. Thus we have
\begin{align}
    -\oint\limits_{C}&\frac{dz}{2\pi i}f(z)G_{a^{\dagger}_{\downarrow}a_{\downarrow}}(\boldsymbol{k},z)G_{a^{\dagger}_{\uparrow}a_{\uparrow}}(\boldsymbol{k}+\boldsymbol{q},z+i\Omega_n)\nonumber\\
    &=\int\limits_{-\infty}^{\infty}\frac{d\epsilon}{2\pi i}f(\epsilon)G_{a^{\dagger}_{\uparrow}a_{\uparrow}}(\boldsymbol{k}+\boldsymbol{q},\epsilon+i\Omega_n)\nonumber\\
    &\times\left[G_{a^{\dagger}_{\downarrow}a_{\downarrow}}(\boldsymbol{k},\epsilon+i0^+)-G_{a^{\dagger}_{\downarrow}a_{\downarrow}}(\boldsymbol{k},\epsilon-i0^+)\right]\nonumber\\
    &+\int\limits_{-\infty}^{\infty}\frac{d\epsilon}{2\pi i}f(\epsilon)G_{a^{\dagger}_{\downarrow}a_{\downarrow}}(\boldsymbol{k},\epsilon-i\Omega_n)\nonumber\\
    &\times\left[G_{a^{\dagger}_{\uparrow}a_{\uparrow}}(\boldsymbol{k}+\vec{q},\epsilon+i0^+)-G_{a^{\dagger}_{\uparrow}a_{\uparrow}}(\boldsymbol{k}+\vec{q},\epsilon-i0^+)\right].
    \label{Integral}
\end{align}

Now we can move on to retarded functions by making the substitution: $i\Omega_n=\Omega+i0^+$ and use the following relation for Green's functions
\begin{align}
	G_{a^{\dagger}_{\sigma}a_{\sigma}}&(\boldsymbol{k},\epsilon+i0^+)-G_{a^{\dagger}_{\sigma}a_{\sigma}}(\boldsymbol{k},\epsilon-i0^+)\nonumber\\
    &=2i\text{Im}[G_{a^{\dagger}_{\sigma}a_{\sigma}}(\boldsymbol{k},\epsilon+i0^+)].
\end{align}
In turn, for the imaginary parts of the Green's functions \eqref{Green} we have
\begin{subequations}
\label{ImGreen}
    \begin{empheq}[]{align}
        \text{Im}&[G_{a^{\dagger}_{\sigma}a_{\sigma}}(\boldsymbol{k},\omega+i0^+)]\nonumber\\
        &=\pi\left[u_{\vec{k},\sigma}\delta(\omega-E^+_{\boldsymbol{k}})+v_{\vec{k},\sigma}\delta(\omega-E^-_{\boldsymbol{k}})\right],\\
        \text{Im}&[G_{b^{\dagger}_{\sigma}b_{\sigma}}(\boldsymbol{k},\omega+i0^+)]\nonumber\\
        &=\pi\left[v_{\vec{k},\sigma}\delta(\omega-E^+_{\boldsymbol{k}})+u_{\vec{k},\sigma}\delta(\omega-E^-_{\boldsymbol{k}})\right],\\
        \text{Im}&[G_{a^{\dagger}_{\sigma}b_{\sigma}}(\boldsymbol{k},\omega+i0^+)]\nonumber\\
        &=\pi w_{\vec{k}}\left[\delta(\omega-E^-_{\boldsymbol{k}})-\delta(\omega-E^+_{\boldsymbol{k}})\right],\\
        \text{Im}&[G_{b^{\dagger}_{\sigma}a_{\sigma}}(\boldsymbol{k},\omega+i0^+)]\nonumber\\
        &=\pi x_{\vec{k}}\left[\delta(\omega-E^-_{\boldsymbol{k}})-\delta(\omega-E^+_{\boldsymbol{k}})\right].
    \end{empheq}
\end{subequations}
Finally we can substitute Eqs.~\eqref{Green} and \eqref{ImGreen} into Eq.~\eqref{Integral} and easily integrate expressions with delta functions. After that we obtain
\begin{align}
    -\oint\limits_{C}&\frac{dz}{2\pi i}f(z)G_{a^{\dagger}_{\downarrow}a_{\downarrow}}(\boldsymbol{k},z)G_{a^{\dagger}_{\uparrow}a_{\uparrow}}(\boldsymbol{k}+\boldsymbol{q},z+i\Omega_n)\nonumber\\
    &=u_{\vec{k+q},\uparrow}u_{\vec{k},\downarrow}\frac{f(E^+_{\vec{k}})-f(E^+_{\vec{k+q}})}{\Omega+i0^+-E^+_{\vec{k+q}}+E^+_{\vec{k}}}\nonumber\\
    &+v_{\vec{k+q},\uparrow}v_{\vec{k},\downarrow}\frac{f(E^-_{\vec{k}})-f(E^-_{\vec{k+q}})}{\Omega+i0^+-E^-_{\vec{k+q}}+E^-_{\vec{k}}}\nonumber\\
    &+u_{\vec{k+q},\uparrow}v_{\vec{k},\downarrow}\frac{f(E^-_{\vec{k}})-f(E^+_{\vec{k+q}})}{\Omega+i0^+-E^+_{\vec{k+q}}+E^-_{\vec{k}}}\nonumber\\
    &+v_{\vec{k+q},\uparrow}u_{\vec{k},\downarrow}\frac{f(E^+_{\vec{k}})-f(E^-_{\vec{k+q}})}{\Omega+i0^+-E^-_{\vec{k+q}}+E^+_{\vec{k}}}.
\end{align}
Performing similar calculations for the remaining bubble diagrams in Eqs.~\eqref{chi_0+-G} and \eqref{chi_0zzG}, after straightforward simplification using expressions \eqref{u_w} and \eqref{v_x}, we obtain Eqs.~\eqref{chi_0+-} and \eqref{chi_0zz}.

\section{Partition function in the bi-orthogonal basis}
In this appendix, we show that using bi-orthogonal basis leads to to the same mean-field equation as previous calculations using the Green's function.

Although the matrix Hamiltonian \eqref{Maxtix_H_MF} is non-Hermitian, it is pseudo-Hermitian, 
\begin{eqnarray}
	\hat{\eta}(\boldsymbol{k}) \hat{H}_{\text{MF}}(\boldsymbol{k})=\hat{H}^{\dagger}_{\text{MF}}(\boldsymbol{k})\hat{\eta}(\boldsymbol{k}),
\end{eqnarray}
\begin{eqnarray}
    \hat{\eta}(\boldsymbol{k})=\begin{bmatrix}
	\cos k_x/\cos k_y &0\\
	0 & 1
	\end{bmatrix}.
\end{eqnarray}
The bi-orthogonal basis includes the following set of right and left eigenvectors
\begin{small}
\begin{subequations}
\label{Basis}
    \begin{empheq}[]{align} 
       &\ket{R^{\pm}_{\vec{k},\sigma}}=N^{\pm}_{\vec{k},\sigma}
	\begin{bmatrix}
	{\left(-\sigma W\mp \sqrt{\varepsilon^2_{\vec{k}}+W^2}\right)}/{t\cos k_x}\\
	1
	\end{bmatrix},\label{R}\\
       &\ket{L^{\pm}_{\vec{k},\sigma}}=N^{\pm}_{\vec{k},\sigma}
	\begin{bmatrix}
	{\left(-\sigma W\mp \left(\sqrt{\varepsilon^2_{\vec{k}}+W^2}\right)^*\right)}/{t\cos k_y}\\
	1
	\end{bmatrix},\label{L}
    \end{empheq}
\end{subequations}
\end{small}
and
\begin{eqnarray}
	\label{N}N^{\pm}_{\vec{k},\sigma}=1/\sqrt{1+\left(\mp\sigma W+\sqrt{\varepsilon^2_{\vec{k}}+W^2}\right)/\varepsilon^2_{\vec{k}}}.
\end{eqnarray}
One can verify that
\begin{eqnarray}
	\sum_{\alpha=\pm}\ket{L^{\alpha}_{\vec{k},\sigma}}\bra{L^{\alpha}_{\vec{k},\sigma}}=\hat{\eta}, \quad \sum_{\alpha=\pm}\ket{R^{\alpha}_{\vec{k},\sigma}}\bra{L^{\alpha}_{\vec{k},\sigma}}=\mathbbm{1}
\end{eqnarray}
and
\begin{eqnarray}
    \bra{L^{\alpha}_{\vec{k},\sigma}}\ket{R^{\beta}_{\vec{k},\sigma}}=\delta_{\alpha\beta},
\end{eqnarray}
where $\alpha,\beta=\pm$.
Calculating the partition function in terms of vectors \eqref{R} and \eqref{L} leads to the standard expression,
\begin{eqnarray}
	Z=e^{-\beta V\frac{W^2}{2U}}\prod_{\boldsymbol{k}}{}^{'}\left(1+e^{-\beta E^{+}_{\boldsymbol{k}}}\right)\left(1+e^{-\beta E^{-}_{\boldsymbol{k}}}\right),
\end{eqnarray}
which we use to obtain the free energy density,
\begin{eqnarray}
	f=-\frac{1}{\beta V}\ln{Z}.
\end{eqnarray}
Now from the condition of free energy density minimization $\partial f/\partial W=0$ we obtain self-consistent equation that completely coincides with Eq.~\eqref{Self_consistent_Eq}.

\bibliography{paper}% Produces the bibliography via BibTeX.

\end{document}